\documentclass[aps,pra,twocolumn,showpacs,floatfix]{revtex4}

\usepackage{graphicx}
\usepackage{nicefrac}
\usepackage{amsmath}
\usepackage{amsfonts}
\usepackage{amssymb}
\usepackage{epsf}
\usepackage{bm}
\usepackage{bbm}
\usepackage{longtable}

\usepackage{dcolumn}

\newcolumntype{.}{D{x}{}{-1}}

\newcommand{\vare}{\varepsilon}
\newcommand{\Za}{Z\alpha}

\newcommand{\bsigma}{\bm{\sigma}}
\newcommand{\balpha}{\bm{\alpha}}

\newcommand{\bgamma}{\bm{\gamma}}

\newcommand{\bfr}{\bm{r}}

\newcommand{\bfp}{\bm{p}}
\newcommand{\bfq}{\bm{q}}

\newcommand{\bfx}{\bm{x}}

\newcommand{\bfz}{\bm{z}}

\newcommand{\hx}{\hat{\bfx}}
\newcommand{\hr}{\hat{\bfr}}
\newcommand{\hz}{\hat{\bfz}}
\newcommand{\hp}{\hat{\bfp}}
\newcommand{\hq}{\hat{\bfq}}

\newcommand{\lbr}{\left<}
\newcommand{\rbr}{\right>}

\newcommand{\pr}{^{\prime}}

\newcommand{\SixJ}[6]{
        \left\{
        \begin{array}{ccc}
        #1  & #2  & #3 \\
        #4  & #5  & #6 \\
        \end{array}
        \right\}
        }
\newcommand{\NineJ}[9]{
        \left\{
        \begin{array}{ccc}
        #1  & #2  & #3 \\
        #4  & #5  & #6 \\
        #7  & #8  & #9 \\
        \end{array}
        \right\}
        }

\newcommand{\intinf}{\int^{\infty}_{-\infty}}

\begin{document}

\title{The two-loop self-energy: diagrams in the coordinate-momentum representation}

\author{Vladimir A. Yerokhin}
\affiliation{Center for Advanced Studies, St.~Petersburg State
Polytechnical University, Polytekhnicheskaya 29,
St.~Petersburg 195251, Russia}

\begin{abstract}
The paper reports a technique of evaluation of Feynman diagrams in the mixed
coordinate-momentum representation. The technique is employed for a
recalculation of the two-loop self-energy correction for the ground state of
hydrogen-like ions with the nuclear charge numbers $Z=10-30$.
The numerical accuracy is considerably improved as compared to the previous
calculations.
The higher-order (in $\Za$) remainder function is inferred from the numerical
results and extrapolated towards $Z=0$ and $1$.
The extrapolated value for hydrogen is consistent (but still not in perfect
agreement) with the analytical result obtained within the perturbative approach.
\end{abstract}


\maketitle

\section{Introduction}

Investigations of the Lamb shift in atomic systems provide one of the most
stringent tests of quantum electrodynamics (QED). They are also used for
the determination of fundamental physical constants \cite{mohr:08:rmp}.
The main factors limiting the present theoretical understanding of the
Lamb shift are the binding two-loop QED effects and, first of all, the two-loop
self-energy correction.

Theoretical investigations of QED effects in light atoms traditionally rely
on the perturbative expansion in the binding-strength parameter $\Za$ ($Z$ is
the nuclear charge and $\alpha$ is the fine structure constant). The
state of the art of such calculations is the evaluation of the dominant part
of the two-loop correction to order $m\alpha^2(\Za)^6$
\cite{pachucki:01:pra,pachucki:03:prl,czarnecki:05:prl,jentschura:05:sese}.
The main problem of the $\Za$-expansion approach is the difficulty of
estimation of uncalculated higher-order effects. In the case of the two-loop
self-energy correction, the higher-order binding effects are above the experimental
error both for the light systems (particularly, for the hydrogen atom
\cite{fischer:04}) and for the heavy ions
\cite{draganic:03,brandau:04,beiersdorfer:05,epp:prl:07}.

In the present investigation, we use the all-order approach which is
nonperturbative in the parameter $\Za$. The nonperturbative calculations started with
the pioneering works of Wichmann and Kroll \cite{wichmann:56} and P.~J.~Mohr 
\cite{mohr:74:a,mohr:74:b}.
For heavy ions, the all-order approach is the only alternative as the parameter of
$\Za$ is of order of unity. For light systems, this method is complementary to
the $\Za$-expansion approach and can provide results for the high-order remainder
beyond the known $\Za$-expansion terms.

The all-order calculation of the two-loop self-energy correction
depicted on Fig.~\ref{fig:sese} was a long and difficult project
accomplished in a series of papers
\cite{mallampalli:98:pra,yerokhin:01:sese,yerokhin:03:prl,yerokhin:03:epjd,%
yerokhin:05:sese,yerokhin:06:prl}.
The numerical results obtained in these studies agreed well with the
first terms of the $\Za$ expansion calculated within the
perturbative approach. However, a significant disagreement was
reported \cite{yerokhin:05:sese}  for the contribution to order
$m\alpha^2(\Za)^6$ (the so-called $B_{60}$ coefficient). A reliable
determination of this contribution from the all-order results
requires a high numerical accuracy to be achieved
for the low values of $Z$, which is a challenging task. In our recent
investigation \cite{yerokhin:09:sese},
we briefly reported a new calculational technique for the evaluation
of Feynman diagrams in the mixed coordinate-momentum representation. This
technique significantly improved the numerical accuracy
of the two-loop self-energy calculation and
to a large extent removed the disagreement with the analytical approach.
In the present paper, we present a detailed description of this calculational
technique.

The relativistic units ($m = \hbar=c=1$) and
the Heaviside charge units ($ \alpha = e^2/4\pi$, $e<0$) will be used throughout
the paper.

\section{Two-loop self-energy}

\begin{figure}[thb]
\centerline{\includegraphics[width=\columnwidth]{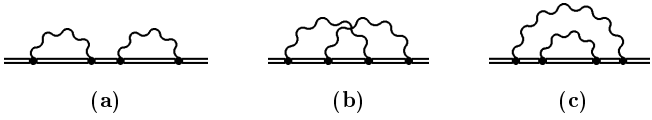}}
\caption{
Feynman diagrams representing the two-loop self-energy correction in the
external binding field. The double line represents the electron propagating
in the field of the nucleus. \label{fig:sese} }
\end{figure}

The Feynman diagrams representing the two-loop self-energy correction are shown
in Fig.~\ref{fig:sese}. The contribution of the first diagram [Fig.~\ref{fig:sese}(a)]
is conveniently divided into two parts, the irreducible and
the reducible one. The reducible part is induced by the virtual states
with the energy $\vare_n=\vare_a$ in the middle electron propagator ($\vare_a$
is the energy of the reference state), and the irreducible part is the
remainder. The irreducible part (often referred to as the loop-after-loop
correction) can be interpreted as a second-order perturbation induced
by the one-loop self-energy operator. The corresponding expression reads
\begin{align}    \label{LAL}
\Delta E_{\rm LAL} = \lbr \gamma^0 \widetilde{\Sigma}(\vare_a)\, G^{\rm
red}\, \gamma^0     \widetilde{\Sigma}(\vare_a)  \rbr\,,
\end{align}
where $\widetilde{\Sigma}(\vare_a)= \Sigma(\vare_a) -\delta m$,
$\Sigma(\vare)$ is the one-loop self-energy operator
\cite{yerokhin:03:epjd}, $\delta
m$ is the corresponding mass counterterm, and $G^{\rm red}$ is the reduced
Dirac-Coulomb Green function. The irreducible part is
finite and can be calculated separately by generalizing various methods
developed for the one-loop self-energy.

%
%

The reducible part is given by
\begin{align}    \label{red}
\Delta E_{\rm red} = \Delta E_{\rm SE}\,\, \Bigl< \left. \gamma^0
    \frac{\partial}{\partial \vare} \widetilde{\Sigma} (\vare)
            \Bigr> \right|_{\vare=\vare_a} ,
\end{align}
where $\Delta E_{\rm SE} = \bigl< \gamma^0\widetilde{\Sigma}(\vare_a)
\bigr>$ is the
one-loop self-energy correction.

The contribution induced by the diagram in Fig.~\ref{fig:sese}(b) will be
referred to as the {\it overlapping} term. It is given by
\begin{align} \label{overlap}
\Delta E_{O} &\ = 2i\alpha
   \intinf d\omega_1\, \int d\bfx_1\ldots d\bfx_4\,
    D(\omega_1,x_{13})\,
   {\psi}^{\dag}_a(\bfx_1)\,
\nonumber \\ &  \times
\alpha_{\mu}\, G(\vare_a-\omega_1)\, \gamma^0
   \Lambda^{\mu}
     (\vare_a-\omega_1,\vare_a)\, \psi_a(\bfx_4)
      - \delta m_O\,,
\end{align}
where $\delta m_O$ is the mass counterterm,
$D(\omega,x_{12})$ is the radial part of the photon propagator in the Feynman
gauge,
\begin{equation}
D(\omega,x_{12}) = \frac{\exp(i\, \sqrt{\omega^2+i0}\,x_{12})}{4\pi x_{12}}
\,,
\end{equation}
$G(\vare)$ is the Dirac-Coulomb Green function defined by $G(\vare) =
[\vare-\mathcal{H}(1-i0)]^{-1}$, with $\mathcal{H}$ being the Dirac
Coulomb Hamiltonian,
and $x_{12} = |\bfx_1-\bfx_2|$. The vertex function $\Lambda^{\mu}$ is defined as
\begin{align}  \label{vertex}
&\displaystyle \Lambda^{\mu}(\vare_a-\omega_1,\vare_a) =
   2i\alpha\gamma^0 \intinf d\omega_2\,
     D(\omega_2,x_{24})\,      \alpha_{\nu}\,
\nonumber \\ &  \times
  G(\vare_a-\omega_1-\omega_2)\,\alpha^{\mu}\, G(\vare_a-\omega_2)\,
         \alpha^{\nu}\,.
\end{align}

The contribution induced by the diagram in Fig.~\ref{fig:sese}(c) will be
referred to as the {\it nested} term. It reads
\begin{align}    \label{nested}
& \Delta E_{N}  = 2i\alpha \intinf d\omega_1\,
   \int d\bfx_1 \ldots d\bfx_4\,
    D(\omega_1,x_{14})\,
       {\psi}^{\dag}_a(\bfx_1)\,
\nonumber \\ & \times
\alpha_{\mu}
          G(\vare_a-\omega_1)\, \gamma^0
           \widetilde{\Sigma}(\vare_a-\omega_1)\,
          G(\vare_a-\omega_1)\,\alpha^{\mu}\, \psi_a(\bfx_4)
\nonumber \\ &
 - \delta m_N\,,
\end{align}
where $\delta m_N$ denotes the mass counterterm.

The general analysis \cite{yerokhin:03:epjd} shows that the sum of the
reducible, the overlapping, and the nested terms is finite. However,
the individual contributions are divergent both in the ultraviolet and the
infrared regions of virtual photon energies. In order to make all
contributions explicitly finite and suitable for a numerical evaluation,
a careful rearrangement of individual parts is required. This rearrangement is
discussed in detail in Ref.~\cite{yerokhin:03:epjd} and will not be repeated
here. The general idea is that the bound electron propagators in the
loops are expanded in terms of the interaction with the binding field and the
resulting contributions are grouped together into three large classes:
(i) the part calculated in the coordinate space, conventionally termed as the
$M$ term and denoted by $\Delta E_{M}$,
(ii) the part calculated in the momentum space (the $F$ term $\Delta E_{F}$), and
(iii) the part calculated in the mixed coordinate-momentum representation (the
$P$ term $\Delta E_{P}$),
\begin{align}
  \Delta E_{\rm red}+  \Delta E_{O} + \Delta E_{N}  =
  \Delta E_{M}+  \Delta E_{F} + \Delta E_{P}\,.
\end{align}
All the three terms can be made explicitly finite and calculated
separately. The calculational technique is completely different for each
term. In the present investigation, we concentrate on the $P$ term, as the
scheme of evaluation of the other two was described in detail in
Ref.~\cite{yerokhin:03:epjd} and has not been changed significantly since
that work.

\section{$\bm{P}$ term: basic formulas}

\begin{figure}
\begin{center}\includegraphics[width=\columnwidth]{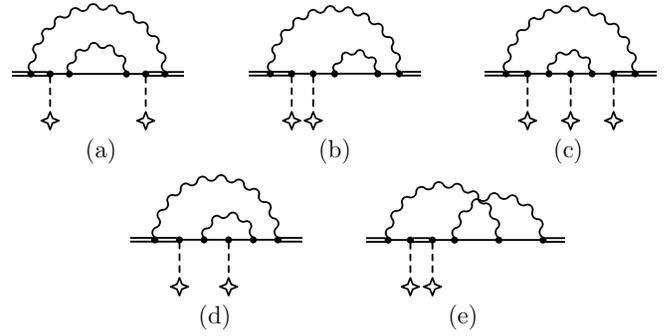}\end{center}%
\caption{\label{fig:pterm}
The $P$ term. The single line represents the free electron propagator. The
dashed line with a cross indicates the interaction with the Coulomb field of
the nucleus.}
\end{figure}

The Feynman diagrams contributing to the $P$ term are shown in
Fig.~\ref{fig:pterm}. They arise from the diagrams in Fig.~\ref{fig:sese}
when the bound-electron propagators in the loops
are expanded in terms of the interaction with the binding
field. The characteristic feature of the diagrams on Fig.~\ref{fig:pterm} is
that the ultraviolet divergences in them originate from the one-loop
subgraphs only. The divergent subgraphs are covariantly regularized and calculated in
the momentum space, whereas the remaining part of the diagrams does not need
any regularization and is evaluated in the coordinate space.

It should be mentioned that the necessity of calculation of Feynman diagrams
in the mixed coordinate-momentum representation is a distinctive feature of
the two-loop effects treated to all orders in the parameter $\Za$. Because
of this, the $P$ term has no analog neither in the one-loop calculations nor, to the
best of our knowledge, in any other previous calculations. (All other two-loop
effects evaluated to all orders in $\Za$
were effectively reduced to one-loop contributions, see
Ref.~\cite{yerokhin:08:twoloop} and references therein.)
For the first time the $P$ term was calculated in
Ref.~\cite{yerokhin:01:sese} with help of a finite basis set representation
of the spectrum of the Dirac equation. In the present investigation we report
a different technique based on the analytical representation of the
Green function in terms of the Whittaker functions.

As shown in Fig.~\ref{fig:pterm}, the $P$ term is represented by a sum of five
terms,
\begin{align}
\Delta E_P = \Delta E_{P,a}+ \Delta E_{P,b}+ \Delta E_{P,c}+ \Delta E_{P,d}+
 \Delta E_{P,e}\,,
\end{align}
each of which refers to the corresponding diagram.
In order to make the individual terms finite, we
assume that the one-loop subgraphs are represented by the renormalized
operators and that the infrared reference-state singularities are removed by
the minimal subtractions. It can be explicitly
checked that such definition of the $P$
term is equivalent to the definition of
Ref.~\cite{yerokhin:03:epjd}, so that the present numerical results are
directly comparable with those of the previous work.

The contribution of the diagram in
Fig.~\ref{fig:pterm}(a) is given by
\begin{align}                                           \label{pa}
\Delta E_{P,a} &  =
        2i\alpha \intinf d\omega\,
                \int \frac{d\bfp}{(2\pi)^3}\,
                        \int d\bfx_1 d \bfx_2 \,
        D(\omega,x_{12})\,
                \nonumber \\ & \times
                 \psi^{\dag}_a(\bfx_1)\, \alpha_{\mu}
      \Bigl[
G_V(E,\bfx_1,\bfp)\,{\mathcal S}_1(E,\bfp)\,
                        G_V(E,\bfp,\bfx_2)
                \nonumber \\ &
-G_V^{(a)}(E,\bfx_1,\bfp)\,{\mathcal S}_1(\vare_a,\bfp)\,
                        G_V^{(a)}(E,\bfp,\bfx_2)
                          \Bigr]
         \alpha^{\mu}        \psi_a(\bfx_2)\, ,
\end{align}
where $E = \vare_a-\omega$,
\begin{equation}
{\mathcal S}_1(\vare,\bfp) = \frac1{\gamma^0\vare-\bgamma\cdot\bfp-m}\,
 \Sigma_R^{(0)}(\vare,\bfp)\,
\frac1{\gamma^0\vare-\bgamma\cdot\bfp-m}
\,,
\end{equation}
$\Sigma_R^{(0)}$ is the renormalized free self-energy operator
(see Ref.~\cite{yerokhin:03:epjd} for its definition and explicit representations),
and the function $G_V$ is the Fourier transform of the product of the
Green function $G$ and the Coulomb potential $V_C(x) = -\Za/x$,
\begin{equation} \label{eq3}
G_V(\vare,\bfx_1,\bfp) =
        \int d\bfx_2 \, e^{i\bfp\cdot \bfx_2} \,
       G(\vare,\bfx_1,\bfx_2) \, V_C(\bfx_2)
         \,,
\end{equation}
\begin{equation} \label{eq4}
G_V(\vare,\bfp,\bfx_2) =
        \int d\bfx_1 \, e^{-i\bfp\cdot \bfx_1}\,
    V_C(\bfx_1)\,
          G(\vare,\bfx_1,\bfx_2) \, .
\end{equation}
The second term in the brackets of Eq.~(\ref{pa}) removes the reference-state
singularity present in the first term.
The definition of the function $G_V^{(a)}$ is obtained from Eqs.~(\ref{eq3})
and (\ref{eq4}) by the substitution $G\to G^{(a)}$, where
$G^{(a)}$ is the reference-state part of the electron propagator defined by
\begin{equation}   \label{Ga}
G^{(a)}(\vare,\bfx_1,\bfx_2) =
     \sum_{\mu_{a'}} \frac{\psi_{a'}(\bfx_1)\,\,
             {\psi}^{\dag}_{a'}(\bfx_2)}{\vare-\vare_a+i0}\,.
\end{equation}
Here, $\psi_{a'}$ denotes the virtual electron state of the same
energy and of the same parity as the reference state and
$\mu_{a'}$ is its momentum projection.

The contribution of the diagram in
Fig.~\ref{fig:pterm}(b) is given by (with the combinatorial factor of 2)
\begin{align}                                           \label{pb}
\Delta E_{P,b} &\  =
        4i\alpha \intinf d\omega\,
                \int \frac{d\bfp}{(2\pi)^3}\,
                        \int d\bfx_1 d \bfx_2 \,
        D(\omega,x_{12})\,
                \nonumber \\ & \times
                 \psi^{\dag}_a(\bfx_1)\, \alpha_{\mu}
\Bigl[G_V(E,\bfx_1,\bfp)-G_V^{(0)}(E,\bfx_1,\bfp)\Bigr]
                \nonumber \\ & \times
  {\mathcal S}_2(E,\bfp)\,
    G^{(0)}(E,\bfp,\bfx_2)\,
         \alpha^{\mu}\,        \psi_a(\bfx_2)\, ,
\end{align}
where
\begin{equation}
{\mathcal S}_2(\vare,\bfp) =
\frac1{\gamma^0\vare-\bgamma\cdot\bfp-m}\,\Sigma_R^{(0)}(\vare,\bfp)\,,
\end{equation}
$G^{(0)}$ is the free Dirac Green function,
and the function $G_V^{(0)}$ is defined by Eqs.~(\ref{eq3})
and (\ref{eq4}) after the substitution $G\to G^{(0)}$.

The contribution of the diagram in
Fig.~\ref{fig:pterm}(c) is
\begin{align}                                           \label{pc}
\Delta E_{P,c} &\  =
        2i\alpha \intinf d\omega\,
                \int \frac{d\bfp_1}{(2\pi)^3}\,\frac{d\bfp_2}{(2\pi)^3}\,
                        \int d\bfx_1 d \bfx_2 \,
                \nonumber \\ & \times
        D(\omega,x_{12})\,V_C(\bfq)\,
                 \psi^{\dag}_a(\bfx_1) \alpha_{\mu}
                \nonumber \\ & \times
      \Bigl[
G_V(E,\bfx_1,\bfp_1)\,{\mathcal G}_1(E,\bfp_1,\bfp_2)\,
                        G_V(E,\bfp_2,\bfx_2)
                \nonumber \\ &
-G_V^{(a)}(E,\bfx_1,\bfp_1)\,{\mathcal G}_1(\vare_a,\bfp_1,\bfp_2)\,
                        G_V^{(a)}(E,\bfp_2,\bfx_2)
                          \Bigr]
                \nonumber \\ & \times
         \alpha^{\mu}\,\psi_a(\bfx_2)\, ,
\end{align}
where $V_C(\bfq) = -4\pi\Za/\bfq^2$ is the Coulomb
potential in momentum space, $\bfq = \bfp_1-\bfp_2$,
\begin{align}
{\mathcal G}_1(\vare,\bfp_1,\bfp_2) = &\
\frac1{\gamma^0\vare-\bgamma\cdot\bfp_1-m}\,
\nonumber \\
& \times
  \Gamma_R^{0}(\vare,\bfp_1;\vare,\bfp_2)\,
\frac1{\gamma^0\vare-\bgamma\cdot\bfp_2-m}\,,
\end{align}
and $\Gamma_R^{0}$ is
the time component of the renormalized
free vertex operator $\Gamma^{\mu}_R$ (its explicit
representation can be found in Ref.~\cite{yerokhin:03:epjd}).

The contribution of the diagram in
Fig.~\ref{fig:pterm}(d) is given by
(with the combinatorial factor of 2)
\begin{align}                                           \label{pd}
\Delta E_{P,d} &\  =
        4i\alpha \intinf d\omega\,
                \int \frac{d\bfp_1}{(2\pi)^3}\,\frac{d\bfp_2}{(2\pi)^3}\,
                        \int d\bfx_1 d \bfx_2 \,
                \nonumber \\ & \times
        D(\omega,x_{12})\,V_C(\bfq)\,
                 \psi^{\dag}_a(\bfx_1)\, \alpha_{\mu}\,
G_V(E,\bfx_1,\bfp_1)
                \nonumber \\ & \times
  {\mathcal G}_2(E,\bfp_1,\bfp_2)\,
    G^{(0)}(E,\bfp_2,\bfx_2)\,
         \alpha^{\mu}\,        \psi_a(\bfx_2)\, ,
\end{align}
where
\begin{align}
{\mathcal G}_2(\vare,\bfp_1,\bfp_2) =
\frac1{\gamma^0\vare-\bgamma\cdot\bfp_1-m}\,
  \Gamma_R^{0}(\vare,\bfp_1;\vare,\bfp_2)\,.
\end{align}

The contribution of the diagram in Fig.~\ref{fig:pterm}(e) can
be written as (with the combinatorial factor of 2)
\begin{align}                                           \label{pe}
\Delta E_{P,e} &\ = -4i\alpha \intinf d\omega\,
\int \frac{d\bfp_1}{(2\pi)^3}\,
        \frac{d\bfp_2}{(2\pi)^3}\,
        \int d\bfz \,
                \nonumber \\ & \times
                \frac{\exp(-i\bfq\cdot \bfz)}{\omega^2 -\bfq^2+ i0}\,
\psi_a^{\dag}(\bfz)\, \alpha_{\mu}\,
                \nonumber \\ & \times
 \Bigl[G_V(E,\bfz,\bfp_1)
-G^{(0)}_V(E,\bfz,\bfp_1)\Bigr]\,
                \nonumber \\ & \times
\frac1{\gamma^0 E-\bgamma\cdot\bfp_1-m}\,
        \Gamma_R^{\mu}(E,\bfp_1;\vare_a,\bfp_2)\,
                        \psi_a(\bfp_2)
\,.
\end{align}

\section{Calculation}

The main problem of the evaluation of the $P$ term is connected with the fact
that the regularization of ultraviolet divergences in the one-loop subgraphs
is carried out in the momentum space, while the bound-electron propagators are
most easily evaluated in the coordinate space. As a result, the expressions
listed in the previous section
contain the Fourier transform of the product of the Dirac Coulomb
Green functions $G$ with the Coulomb potential $V_C$ over one of the radial
arguments, see Eqs.~(\ref{eq3}) and (\ref{eq4}). Since we were not able to
find a satisfactory analytical representation for such an object, the only
possible way was to evaluate the Fourier tansform integral numerically.
This way entails
numerous numerical integrations of rapidly oscillating functions, which may make
the computations prohibitively expensive, unless great care is taken in choosing the
optimal calculational approach.

In the previous investigations \cite{yerokhin:01:sese,yerokhin:03:prl,yerokhin:03:epjd},
the $P$ term was calculated with help of a finite basis set
for the Dirac equation. The advantage of the basis-set
methods is that they represent the Dirac Green function as a continuous
function of the radial arguments (for any finite size of the basis), whereas the
exact Green function is discontinous when the two radial arguments are equal.
The usage of a finite basis set allows one to perform the Fourier transform
of the Green function over one radial variable independently of the other.
However, the convergence with respect to the size of the basis
appears to be the limiting factor for the accuracy of the
calculations. In the present investigation, we set up a
different calculational approach
based on the analytical representation of the Green function.
Technical details of evaluation of the
Dirac Coulomb Green function in the mixed coordinate-momentum representation
are described in Appendix~\ref{app:DCGF}. The corresponding formulas for the
free Dirac Green function are summarized in Appendix~\ref{app:freeDGF}.

\subsection{Nested diagrams}
\label{subsec:n}

In this subsection, we address the diagrams in Figs.~\ref{fig:pterm}(a)-(d)
and outline the major steps required to make the basic formulas
suitable for a numerical evaluation.

The integrations over the angular variables [$\hx_1$, $\hx_2$, and $\hp$ in
Eqs.~(\ref{pa}) and (\ref{pb}) and $\hx_1$, $\hx_2$, $\hp_1$, and $\hp_2$ in
Eqs.~(\ref{pc}) and (\ref{pd}), where $\hx = \bfx/|\bfx|$]
are relatively simple. We employ the fact that
the matrix elements of the operators ${\mathcal S}_{1,2}$ and ${\mathcal G}_{1,2}$
with the Dirac wave functions are
(i) diagonal with respect to the angular momentum quantum number $\kappa$ and
the momentum projection $\mu$ and (ii) do not depend on $\mu$:
\begin{align}
\lbr \kappa\mu|{\mathcal S}_i|\kappa\pr\mu\pr\rbr &\, =
\delta_{\kappa,\kappa\pr}\,\delta_{\mu,\mu\pr}
\lbr \kappa|{\mathcal S}_i|\kappa\rbr\,,
 \\
\lbr \kappa\mu|{\mathcal G}_i\,V_C|\kappa\pr\mu\pr\rbr &\, =
\delta_{\kappa,\kappa\pr}\,\delta_{\mu,\mu\pr}
\lbr \kappa|{\mathcal G}_i\,V_C|\kappa\rbr\,.
\end{align}
As a result, the integrations over $\hp$ in Eqs.~(\ref{pa}) and (\ref{pb})
and $\hp_1$ and $\hp_2$ in Eqs.~(\ref{pc}) and (\ref{pd}) are exactly the same
as for the zero- and one-potential parts of the one-loop self-energy correction,
see Ref.~\cite{yerokhin:99:pra} for details. The integrations over $\hx_1$ and
$\hx_2$ are the same as for the many-potential part of the one-loop
self-energy correction.

As illustrated in Ref.~\cite{yerokhin:99:pra}, the integations over
$\hp_1$ and $\hp_2$ in Eqs.~(\ref{pc}) and (\ref{pd}) can be easily reduced to
a single integral over $\xi = \hp_1\cdot\hp_2$, which needs to be evaluated
numerically. The calculation is complicated by the presence of an integrable
Coulomb singularity ($\sim 1/q^2$, $q = |\bfp_1-\bfp_2|$). This singularity is
removed in two steps. First, the change of the integration variable
$\xi\to q$ weakens it to $\sim 1/q$. The remaining singularity
is removed by subtraction of the vertex function with the equal arguments,
\begin{equation}
  \Gamma_R^{0}(p_1,p_2) \to
  \Gamma_R^{0}(p_1,p_2)-\frac12\left[\Gamma_R^{0}(p_1,p_1)
+\Gamma_R^{0}(p_2,p_2)\right]\,.
\end{equation}
The vertex operator with two equal arguments is related to the free self-energy
operator by the Ward identity:
\begin{equation}
  \Gamma_R^{0}(p,p) = -\frac{\partial}{\partial p^0}\,\Sigma^{(0)}_R(p)\,.
\end{equation}
In the subtracted terms, the Coulomb singularity is easily integrated out
by using identities obtained from
the definition of the Dirac Coulomb Green function,
as,  e.g.,
\begin{align}
\int \frac{d\bfp_2}{(2\pi)^3}\, &\, V_C(\bfq)\,
     \frac1{\gamma^0E-\bgamma\cdot\bfp_2-m}\, G_V(E,\bfp_2,\bfx_2) =
 \nonumber \\
 & G_V(E,\bfp_1,\bfx_2)-G_V^{(0)}(E,\bfp_1,\bfx_2)\,.
\end{align}

Finally, we change the contour of the integration over the virtual photon
energy $\omega$ in Eqs.~(\ref{pa}), (\ref{pb}), (\ref{pc}), and (\ref{pd})
from $(-\infty,\infty)$ to a new contour $C_{LH}$, whose main part is
parallel to the imaginary axis. The contour $C_{LH}$
consists of the low-energy $C_L$ and the high-energy $C_H$ parts.
The low-energy part extends over $(\Delta-i0,-i0)$ on
the lower bank of the cut of the photon propagator and over $(i0,\Delta+i0)$ on
the upper bank of the cut, with the parameter $\Delta$ fixed by
$\Delta = \Za\, \vare_a$. The high-energy part
consists of two intervals, $(\Delta+i0,\Delta+i\infty)$ and
$(\Delta-i0,\Delta-i\infty)$. The contour $C_{LH}$ differs from
the one used by P.~J.~Mohr \cite{mohr:74:a} only by the choice of the breaking
point $\Delta$ (the value $\Delta = \vare_a$ was employed in that work).

In our previous calculations we used the contour that extended along the
imaginary axis (which corresponds to the choice of $\Delta=0$). The contour
$C_{LH}$ is more convenient for the numerical evaluation. Firstly,
there is no pole contributions originating from the reference-state part of
the electron propagators.
Secondly and more importantly, the photon propagator on the low-energy part
of the contour involves $\sin (\omega x_{12})$, which suppresses
small denominators due to the virtual bound states
and leads to a smooth behaviour of the integrand for small $\omega$.

In the present investigation, we are concerned with the reference
state being the ground state only. In this case, no pole contributions appears
for the contour $C_{LH}$. For the excited reference states, however, there are
single and double poles on the low-energy part of the contour, which arise from
the intermediate states bounded more deeply than the reference state. These poles
require a separate treatment or a deformation of the integration contour
into the complex plane.

\subsection{Overlapping diagram}
\label{subsec:o}

In this subsection we address the overlapping diagram shown in
Fig.~\ref{fig:pterm}(e), whose expression is given by Eq.~(\ref{pe}).
The angular integration in this expression
is rather involved and will be considered in detail.

In order to perform the integration over $\hz$, we expand the exponent
into the spherical waves,
\begin{equation}
  e^{-i \bfq\cdot \bfz} = 4\pi \sum_{LM} i^{-L} j_L(qz)\,
  Y_{LM}(\hq)\, Y_{LM}^*(\hz)\,,
\end{equation}
where $j_L$ is the spherical Bessel function and $Y_{LM}$ is the spherical
harmonics. The time component ($\mu = 0$) of the $\hz$ integration is
immediately evaluated in terms of the basic integrals of the form
\begin{align}
\int d\hz\, \chi^{\dag}_{\kappa_b \mu_b}(\hz)\, Y_{LM}(\hz)\,
        \chi_{\kappa_a \mu_a}(\hz) = s_{LM}^{ba}\, \langle \kappa_b ||\,{\bm
          C}^{(L)}||\kappa_a\rangle \, ,
\end{align}
where $\chi_{\kappa \mu}(\hz)$ are the Dirac spin-angular spinors \cite{rose:61},
${\bm C}^{(L)}$ is the spherical tensor with components ${\bm C}^{(L)}_M(\hr)
= \sqrt{4\pi/(2L+1)}\,Y_{LM}(\hr)$, $\langle  ||\cdots||\rangle$ denotes the
reduced matrix element, and
\begin{align}
 s_{LM}^{ba} = \frac{(-1)^{j_a-\mu_a}}{\sqrt{4\pi}}\, C^{LM}_{j_b\mu_b, j_a -\mu_a}
             \, ,
\end{align}
with $C_{j_1m_1,j_2m_2}^{jm}$ being the Clebsch-Gordan coefficient.

The vector components ($\mu = 1$, 2, 3) of the  $\hz$ integration are calculated
after expanding the integrand in terms of the vector spherical
harmonics ${\bf Y}_{JLM}$ \cite{johnson:88:b,varshalovich},
\begin{align}
\chi^{\dag}_{\kappa_b \mu_b}(\hz)\, \bsigma
        \chi_{\kappa_a \mu_a}(\hz)  =
        \sum_{JLM} s_{JM}^{ab}\, S_{JL}(\kappa_b,\kappa_a)\,
             {\bf Y}_{JLM}(\hz) \, ,
\end{align}
where $\bsigma$ is a vector incorporating Pauli matrices. The
coefficients $S_{JL}$ are given by
\begin{align} \label{SJL}
S_{J\, J+1}(\kappa_a ,\kappa_b ) =&\,
  \sqrt{\frac{J+1}{2J+1}}
  \left( 1+ \frac{\kappa_a +\kappa_b}{J+1} \right)\,
\nonumber \\ & \times
    \langle -\kappa_b ||\,{\bm C}^{(J)}||\kappa_a\rangle\,,
\\
  S_{J\, J}(\kappa_a,\kappa_b ) =&\, \frac{\kappa_a-\kappa_b}{\sqrt{J(J+1)}}\,
    \langle \kappa_b ||\,{\bm C}^{(J)}||\kappa_a\rangle\,,
\\
  S_{J\, J-1}(\kappa_a,\kappa_b ) =&\, \sqrt{\frac{J}{2J+1}}
  \left( -1+ \frac{\kappa_a+\kappa_b}{J} \right) \,
\nonumber \\ & \times
    \langle -\kappa_b ||\,{\bm C}^{(J)}||\kappa_a\rangle\,.
\end{align}
For $J=0$, the only nonvanishing coefficient is $S_{01}$.

We now turn to the evaluation of the integrals over $\hp_1$ and $\hp_2$ in
Eq.~(\ref{pe}). The aim is to integrate out all angular variables except
$\xi = \hp_1\cdot\hp_2$. To this end,
we examine the angular structures encountered in the integrand. The time
component of the vertex operator 
sandwiched between the Dirac wave functions
involves two independent angular structures,
which are identified by
\begin{align} \label{vertex3}
&
\psi^{\dag}_n(\bfp_1) \,
\frac1{\gamma^0 E-\bgamma\cdot\bfp_1-m}\,
        \Gamma_R^{0}(E,\bfp_1;\vare_a,\bfp_2)\,
   \psi_a(\bfp_2)
   \nonumber \\
&    =   \frac{\alpha}{4\pi}\, i^{l_n-l_a} \biggl\{
        \bigl[g_n\,{\mathcal F}_{1g}+f_n\,{\mathcal F}_{1f}\bigr]\,
              \chi^{\dag}_{\kappa_n \mu_n}(\hp_1)\,
        \chi_{\kappa_a \mu_a}(\hp_2)
\nonumber \\
 & +
        \bigl[g_n\,{\mathcal F}_{2g}+f_n\,{\mathcal F}_{2f}\bigr]\,
     \chi^{\dag}_{-\kappa_n \mu_n}(\hp_1)\,
        \chi_{-\kappa_a \mu_a}(\hp_2) \biggr\} \, ,
\end{align}
where $g_n\equiv g_n(p_1)$ and $f_n\equiv f_n(p_1)$ are the upper and the lower radial components of the
Dirac wave function $\psi_n$ and
${\mathcal F}_i$ are scalar functions ${\mathcal F}_i \equiv {\mathcal
  F}_i(E,\vare_a,p_1,p_2,q)$, with $p_1 = |\bfp_1|$, $p_2 = |\bfp_2|$,
and $q = |\bfq|$. The vector part of the vertex operator induces six
angular structures,
\begin{align}
        \label{vertex3a}
& \psi^{\dag}_n(\bfp_1) \,
\frac1{\gamma^0 E-\bgamma\cdot\bfp_1-m}\,
        \bm{\Gamma}_R(E,\bfp_1;\vare_a,\bfp_2)\,
   \psi_a(\bfp_2)
   \nonumber \\
&
        = \frac{\alpha}{4\pi}\, i^{l_n-l_a}\,
\Bigl\{
        \bigl[g_n\,{\mathcal R}_{1g}+f_n\,{\mathcal R}_{1f}\bigr]\,
    \chi^{\dag}_{\kappa_n \mu_n}(\hat{\bfp}_1)
        \,{\bsigma}\, \chi_{-\kappa_a\mu_a}(\hat{\bfp}_2)
\nonumber \\
 &
+        \bigl[g_n\,{\mathcal R}_{2g}+f_n\,{\mathcal R}_{2f}\bigr]\,
    \chi^{\dag}_{-\kappa_n \mu_n}(\hat{\bfp}_1)
        \,{\bsigma}\, \chi_{\kappa_a\mu_a}(\hat{\bfp}_2)
         \nonumber \\
&
+        \bigl[g_n\,{\mathcal R}_{3g}+f_n\,{\mathcal R}_{3f}\bigr]\,
       \bfp_1\,
         \chi^{\dag}_{\kappa_n \mu_n}(\hat{\bfp}_1)
         \chi_{\kappa_a\mu_a}(\hat{\bfp}_2)
         \nonumber \\
&
+        \bigl[g_n\,{\mathcal R}_{4g}+f_n\,{\mathcal R}_{4f}\bigr]\,
      \bfp_2\,
         \chi^{\dag}_{\kappa_n \mu_n}(\hat{\bfp}_1)
         \chi_{\kappa_a\mu_a}(\hat{\bfp}_2)
         \nonumber \\
&
+        \bigl[g_n\,{\mathcal R}_{5g}+f_n\,{\mathcal R}_{5f}\bigr]\,
       \bfp_1\,
         \chi^{\dag}_{-\kappa_n \mu_n}(\hat{\bfp}_1)
         \chi_{-\kappa_a\mu_a}(\hat{\bfp}_2)
         \nonumber \\
&
+        \bigl[g_n\,{\mathcal R}_{6g}+f_n\,{\mathcal R}_{6f}\bigr]\,
      \bfp_2\,
         \chi^{\dag}_{-\kappa_n \mu_n}(\hat{\bfp}_1)
         \chi_{-\kappa_a\mu_a}(\hat{\bfp}_2)
\Bigr\}\,,
\end{align}
where ${\mathcal R}_i \equiv {\mathcal
  R}_i(E,\vare_a,p_1,p_2,q)$. The functions ${\mathcal F}_i$
and ${\mathcal R}_i$ can be straightforwardly obtained from formulas
in Appendix A of Ref.~{\cite{yerokhin:99:sescr}}.

Using Eqs.~(\ref{vertex3}) and (\ref{vertex3a}), it is possible to parameterize
the angular structure of the integrand of Eq.~(\ref{pe}) by four
basic angular factors $t_{\kappa_1,\kappa_2}$,
$s_{\kappa_1,\kappa_2}^{\sigma}$, $s_{\kappa_1,\kappa_2}^{p_1}$, and $s_{\kappa_1,\kappa_2}^{p_2}$
defined as
\begin{align}    \label{pef6}
t_{\kappa_n,\kappa_a}(J) = \sum_{\mu_n M}
    s_{JM}^{na}\,
    \chi^{\dag}_{\kappa_n \mu_n}(\hp_1)\, Y_{JM}(\hq)\,
        \chi_{\kappa_a \mu_a}(\hp_2) \,,
\end{align}
\begin{align}
s_{\kappa_n,\kappa_a}^{\sigma}(JL) = \sum_{\mu_n M}
    s_{JM}^{na}\,
    \chi^{\dag}_{\kappa_n \mu_n}(\hp_1)\, {\bsigma}\cdot{\bf Y}_{JLM}(\hq)\,
        \chi_{\kappa_a \mu_a}(\hp_2) \,,
\end{align}
\begin{align}
s_{\kappa_n,\kappa_a}^{p_i}(JL) = \sum_{\mu_n M}
    s_{JM}^{na}\,
    \chi^{\dag}_{\kappa_n \mu_n}(\hp_1)\, {\hp_i}\cdot{\bf Y}_{JLM}(\hq)\,
        \chi_{\kappa_a \mu_a}(\hp_2) \,.
\label{pef6c}
\end{align}
By an explicit evaluation with help of formulas from Ref.~\cite{varshalovich},
one can show that the above angular factors are
the functions of $p_1$, $p_2$, and $q$ only (or, in other words,
that they depend on the angular variables only through $\xi$). This
statement allows us to integrate out all angles in Eq.~(\ref{pe})
except $\xi$. The calculation of angular factors is described in
Appendix~\ref{app:angular}.

For a numerical evaluation of Eq.~(\ref{pe}), we need to deform the contour
of the $\omega$ integration. In this case (in contrast to the nested
contributions), we find it convenient just to rotate the integration contour
to the imaginary axis, $\omega\to i\omega$. This leads to appearance of the
pole contribution. So,
\begin{equation}
\Delta E_{P,e} = \Delta E_{P,e}^{\rm Im}+ \Delta E_{P,e}^{\rm pole}\,,
\end{equation}
where $\Delta E_{P,e}^{\rm Im}$ is the contribution of the integral along the
imaginary axis and $\Delta E_{P,e}^{\rm pole}$ is the pole contribution.

The final result after the angular integrations and the rotation of the
contour is
\begin{widetext}
\begin{align} \label{pe3}
\Delta E_{P,e}^{\rm Im} &= \frac{\alpha^2}{\pi^4}\,\Re\,\sum_{\kappa_n}
\int_0^{\infty} d\omega\,\,  \int_0^{\infty} dp_1\,dq\,
  \int_{|p_1-q|}^{p_1+q}d p_2\,
   \frac{q\,p_1p_2}{-\omega^2-q^2}\, \int_0^{\infty} dz\, z^2\,
                \nonumber \\ & \times
\Biggl\{ \sum_J  (-1)^{k_1}\, j_J(qz)\,
            \langle \kappa_n||{\bm C}^{(J)}||\kappa_a\rangle\,
     \Bigl[\bigl(g_a\,{\widetilde{G}}_{V_{\kappa_n}}^{11}+f_a\,{\widetilde{G}}_{V_{\kappa_n}}^{21}\bigr)\,{\mathcal F}_g +
           \bigl(g_a\,{\widetilde{G}}_{V_{\kappa_n}}^{12}+f_a\,{\widetilde{G}}_{V_{\kappa_n}}^{22}\bigr)\,{\mathcal F}_f
\Bigr]
                \nonumber \\ &
 - \sum_{JL} (-1)^{k_2}\,  j_L(qz)\,
       \Bigl(\bigl[g_a\,{\widetilde{G}}_{V_{\kappa_n}}^{21}\,S_{JL}(\kappa_a,-\kappa_n)
                   -f_a\,{\widetilde{G}}_{V_{\kappa_n}}^{11}\,S_{JL}(-\kappa_a,\kappa_n)\bigr]
\,{\mathcal R}_g
                                  \nonumber \\ &
+\bigl[g_a\,{\widetilde{G}}_{V_{\kappa_n}}^{22}\,S_{JL}(\kappa_a,-\kappa_n)
                   -f_a\,{\widetilde{G}}_{V_{\kappa_n}}^{12}\,S_{JL}(-\kappa_a,\kappa_n)\bigr]\,
{\mathcal R}_f  \Bigr)   \Biggr\}
 \, ,
\end{align}
\end{widetext}
where $k_1 = (J-l_n+l_a)/2$, $k_2 = (L-l_n+l_a-1)/2$,
$l_n = |\kappa_n+1/2|-1/2$,  $g_a \equiv
g_a(z)$ and $f_a \equiv f_a(z)$ are the radial components of the
reference-state wave function, and ${\widetilde{G}}_{V_{\kappa_n}}^{ij}$ stand for the difference
of the radial components of the Coulomb Green function (times $V_C$)
and the free Green function (times $V_C$),
$$
{\widetilde{G}}_{V_{\kappa_n}}^{ij}\equiv G^{ij}_{V_{\kappa_n}}(\vare_a-i\omega,z,p_1)
  - G^{{(0)}^{ij}}_{V_{\kappa_n}}(\vare_a-i\omega,z,p_1)\,.
$$
The angular functions in Eq.~(\ref{pe3}) are defined by
\begin{equation}
    {\mathcal F}_g =
    {\mathcal F}_{1g}\,\,t_{\kappa_n,\kappa_a}(J)
 +{\mathcal F}_{2g}\,\,t_{-\kappa_n,-\kappa_a}(J)\,,
\end{equation}
\begin{align}
{\mathcal R}_g & \ =
  {\mathcal R}_{1g}\,s_{\kappa_n,-\kappa_a}^{\sigma}(JL)+{\mathcal R}_{2g}\,\,s_{-\kappa_n,\kappa_a}^{\sigma}(JL)
  \nonumber \\ &
  +p_1{\mathcal R}_{3g}\,s_{\kappa_n,\kappa_a}^{p_1}(JL)+p_2{\mathcal R}_{4g}\,s_{\kappa_n,\kappa_a}^{p_2}(JL)
\nonumber \\ &
+p_1{\mathcal R}_{5g}\,s_{-\kappa_n,-\kappa_a}^{p_1}(JL)
         +p_2{\mathcal R}_{6g}\,s_{-\kappa_n,-\kappa_a}^{p_2}(JL)\,,
\end{align}
and the same for the ${\mathcal F}_f$ and ${\mathcal R}_f$ functions.

The pole contribution $\Delta E_{P,e}^{\rm pole}$ is
obtained from Eq.~(\ref{pe3}) by the following
substitution (valid for $a$ being the ground state),
\begin{equation}
{\widetilde{G}}_{V_{\kappa_n}}^{ij}
  \to -\frac{\pi}{2}\,\delta_{\kappa_n\kappa_a}\,\delta(\omega)\, \phi_{a}^i(z)\,\phi_{Va}^j(p_1)\,,
\end{equation}
where $\phi^1_a(z) = g_a(z)$, $\phi^2_a(z) = f_a(z)$, 
and $\phi_{Va}^i(p)$ is the Fourier transform of the product $\phi_a^i(x)\,V_C(x)$.

A useful check of the angular-momentum algebra
consists in making the substitution $\Gamma_R^{\mu}(E,\bfp_1;\vare_a,\bfp_2) \to
\gamma^{\mu}$ in Eq.~(\ref{pe}). The result is an one-loop self-energy
contribution which can be calculated
independently in the coordinate representation.


\subsection{Numerical evaluation}

We start our discussion of the numerical evaluation of the $P$ term with
$\Delta E_{P,a}$ and $\Delta E_{P,b}$ given by
Eqs.~(\ref{pa}) and (\ref{pb}), respectively. These are the two simplest
contributions. After carrying out integrations
over the angular variables as described in Sec.~{\ref{subsec:n}},
five integrations remain
to be calculated numerically, namely those
over $\omega$, $p$, $x_1$, and $x_2$ and the
Bessel transform integral implicitly present in the Green function.

The radial integrations over $x_1$ and
$x_2$ have to be organized in such a way as to avoid
unnecessary recalculation of the Bessel
transform integrals, as discussed in Appendix~\ref{app:DCGF}. To this end, we set up
a radial grid $\left\{x_{i,j,k}\right\}$ as follows. The
first-level elements $x_{i,0,0}$ are given by
\begin{equation}
 x_{i,0,0} = \rho_0\, \frac{1-t_i^2}{t_i^2}\,,
\end{equation}
where $\rho_0$ is a parameter adjusted empirically,
the variable $t_i$ is uniformly distributed over the interval $(t_{\rm
  min},1)$, and a small value of $t_{\rm min}>0$ cuts off
the radial integrations at large distances. The second-level elements $x_{i,j,0}$
represent the Gauss-Legendre quadratures on the interval
$(x_{i,0,0},x_{i+1,0,0})$. The third-level elements $x_{i,j,k}$ represent the
Gauss-Legendre quadratures on the interval $(x_{i,j,0},x_{i,j+1,0})$.
In the result, we obtain an ordered three-level radial grid. To
perform the radial integrations over $x_1$ and $x_2$, it is sufficient to know
the integrand on this grid only.

The general scheme of the evaluation of $\Delta E_{P,a}$
looks as follows. For fixed values of
$\kappa$, $\omega$, and $p$, we set up the radial grid $\left\{x_{i,j,k}\right\}$. On this
radial grid, we store the components of the Dirac Green function
$\phi_{\kappa}^{0}(E,x)$ and $\phi_{\kappa}^{\infty}(E,x)$ [see Eqs.~(\ref{Green2})
and (\ref{Green2a})],
the Bessel transform functions $\psi_{\kappa}^{0}(E,p;x)$ and
$\psi_{\kappa}^{\infty}(E,p;x)$ [see Eqs.~(\ref{gr3}) and (\ref{gr4})], and the other
functions required for the evaluation of
the integrand (the radial part of the photon propagator, the
reference-state wave function, {\em etc.}). After that, the
radial integrations are performed simply by summing up the stored numerical
values. Next, we perform the integration over $p$, then the one over $\omega$,
and finally, the summation over $\kappa$.

The most expensive part of the calculation is the evaluation of the Bessel
transform integrals. In order to control the accuracy of numerical
integrations, one needs an efficient procedure for calculating
the transforms for various momenta, including ones as large as
$10^6$. In our calculations, we used the Gauss-Legendre integration quadratures in
the region where the argument of the Bessel function is of about unity or
smaller. Outside this region, the spherical Bessel function was expressed as a
combination of the sine and cosine functions. The numerical evaluation of the
sine and cosine transform was performed with help of routines of the
NAG Fortran library.

The scheme described above works well for the $\Delta E_{P,a}$ contribution
but turns out to be not sufficiently effective for
$\Delta E_{P,b}$, leading to a slow convergence of
the radial integrations with respect to the number of integration points. This
is because the free Green function $G^{(0)}(\vare,\bfx,\bfp)$
contains a Bessel function [see Eq.~(\ref{fgr3})], which
oscillates rapidly in the high-momenta region. This problem was solved by
observing that the integral over $x_2$ in Eq.~(\ref{pb}) has a structure
similar to $\psi_{\kappa}(E,p;x_1)$, i.e., it is essentially a Bessel transform over the
intervals $(0,x_1)$ and $(x_1,\infty)$. We thus perform the integration over
$x_2$ in Eq.~(\ref{pb}) by means of the same approach as used in the evaluation
of the Bessel transform functions $\psi_{\kappa}(E,p;x_1)$. This approach
improves the convergence of the radial integrals drastically.

The evaluation of the two remaining nested contributions, $\Delta
E_{P,c}$ and  $\Delta E_{P,d}$, was performed in the full analogy with the discussed
above. However, it turned out to be much more time consuming due to
a larger number of integrations. Indeed, in place of an integration over
$p$ in $\Delta
E_{P,a}$ and $\Delta E_{P,b}$, there are now four integrations (those
over $p_1$, $p_2$, and $q$ and
the Feynman-parameter integration implicitly present in the vertex
operator). Fortunately, the integrations over $q$ and over the Feynman parameter
can be carried out independently of the integrations over $x_1$ and $x_2$ and
thus do not  significantly influence the total calculational time. The two
integrations over $p_1$ and $p_2$, however, lead to a considerable increase of
the computational expense (this being about a week of the processor
time for each value of $Z$).

For the overlapping contribution $\Delta E_{P,e}$, there are seven
integrations to be performed numerically. Five of them are explicitly written
in Eq.~(\ref{pe3}), one over the Feynman parameter is implicitly present in
the vertex operator, and the last one is the Bessel transform integral in the
Green function. The number of nested integrations can be reduced by
observing that the integrations over $p_2$ and over the Feynman parameter can be
carried out independently of the integration over $z$.
The calculation is complicated by the fact that the
integral over $z$ contains a Bessel function, thus being essentially a
Bessel transform. In order to get a stable result for the $z$ integration in
the region of large values of $qz$, we had to interpolate the part of the
integrand that multiplies the Bessel function and evaluate the Bessel
transform analytically.

\section{Results and discussion}

The results of our calculation of the $P$ term for the ground state
of hydrogen-like ions with $Z\ge 10$
are listed in Table~\ref{tab:pterm}. The
individual contributions are presented in a way that allows a detailed comparison with the previous
calculations. Specifically, the first four columns of
Table~\ref{tab:pterm} are directly comparable to the four columns of Table~2 in
Ref.~\cite{yerokhin:03:epjd}. The previous results listed in
the fifth column of Table~\ref{tab:pterm} were obtained in Ref.~\cite{yerokhin:05:sese} for $Z\le
60$ and in Ref.~\cite{yerokhin:03:epjd} for the other $Z$.
The agreement with the previous calculations is
very good in most cases, but the present numerical accuracy is
significantly higher. A small deviation in the high-$Z$ region is probably due
to a difference in the treatment of the nucleus. (In the present work, the point
nuclear model is used, whereas the previous investigations
\cite{yerokhin:05:sese,yerokhin:03:epjd} were conducted with a partial
inclusion of the finite nuclear size effect in the $P$ term.)

%
%
\begin{table*}[htb]
\caption{The $P$ term
for the ground state of hydrogen-like ions, in units of $\Delta
E/[m\alpha^2(\Za)^4/\pi^2]$.
 \label{tab:pterm}}
\begin{center}
\begin{tabular}{r.....}
\hline\noalign{\smallskip}
$Z$  &  \multicolumn{1}{c}{Figs.~(a,b)}
              &  \multicolumn{1}{c}{Figs.~(c,d)}
                             &  \multicolumn{1}{c}{Fig.~(e)}
                                               &  \multicolumn{1}{c}{Total}
                                               &  \multicolumn{1}{c}{Previous
                                                 \cite{yerokhin:05:sese,yerokhin:03:epjd}}
    \\
\noalign{\smallskip}\hline\noalign{\smallskip}
  10 & -855.x4289\,(20)  & 1265.x4550\,(50) & -1131.x3372\,(22)  & -721.x3111\,(58) &-721.x34\,(12)\\
  12 & -486.x0740\,(20)  &  744.x6950\,(50) &  -697.x6864\,(17)  & -439.x0654\,(56) &\\
  15 & -239.x2533\,(15)  &  384.x3590\,(30) &  -380.x3170\,(12)  & -235.x2113\,(36) &-235.x205\,(70)\\
  17 & -159.x4480\,(15)  &  263.x5625\,(20) &  -268.x3945\,(12)  & -164.x2800\,(28) &\\
  20 &  -93.x3597\,(7)   &  160.x3551\,(20) &  -169.x0247\,(10)  & -102.x0293\,(23) &-102.x026\,(55)\\
  25 &  -44.x1892\,(7)   &   79.x9950\,(15) &   -87.x7882\,(10)  &  -51.x9824\,(19) &\\
  30 &  -23.x8155\,(7)   &   44.x8102\,(27) &   -50.x4083\,(10)  &  -29.x4135\,(30) &-29.x410\,(25)\\
  40 &   -9.x0138\,(4)   &   17.x6061\,(2)  &   -20.x1713\,( 6)  &  -11.x5790\,( 7) &-11.x575\,(30)\\
  50 &   -4.x3391\,(5)   &    8.x4020\,(4)  &    -9.x5506\,( 4)  &   -5.x4877\,( 7) &-5.x488\,(26)\\
  60 &   -2.x4455\,(3)   &    4.x5451\,(4)  &    -5.x0660\,( 2)  &   -2.x9664\,( 5) &-2.x970\,(18)\\
  70 &   -1.x5203\,(2)   &    2.x6716\,(1)  &    -2.x9354\,( 2)  &   -1.x7841\,( 3) &-1.x757\,(25)\\
  83 &   -0.x8655\,(2)   &    1.x4268\,(1)  &    -1.x6307\,( 1)  &   -1.x0693\,( 2) &-1.x057\,(13)\\
  92 &   -0.x5545\,(1)   &    0.x9091\,(1)  &    -1.x1902\,( 1)  &   -0.x8356\,( 2) &-0.x812\,(10)\\
 100 &   -0.x2990\,(2)   &    0.x5426\,(1)  &    -0.x9792\,( 4)  &   -0.x7356\,( 5) &-0.x723\,(7)\\
\noalign{\smallskip}\hline
\end{tabular}
\end{center}
\end{table*}

The uncertainty of the present results is mainly due to the termination of the
partial-wave expansion. In our calculation, we included typically 20-30
partial waves and estimated the omitted tail by fitting the data obtained
as a function of the cutoff parameter. Perspectives
for improving the present accuracy further (which is required if one is to perform a
calculation for lower values of $Z$) seem to be questionable. The main problem is
that the high partial waves become increasingly difficult to control
numerically. At the same time, the extrapolation of the partial wave expansion
requires an accurate representation of the individual partial-wave expansion terms.

The main motivation of the present investigation was to improve the numerical accuracy of
the total two-loop self-energy correction in the region of medium values of
$Z$, in order to get a more reliable extrapolation towards $Z=0$. To this
end, a new approach was developed for the evaluation of the $P$ term, as
described above. Besides that, the other parts of the two-loop
self-energy correction were reevaluated to a higher accuracy. This was
accomplished by the methods described in Ref.~\cite{yerokhin:03:epjd}, with
the increased number of partial waves included and with denser integration
grids. The results were first presented in Ref.~\cite{yerokhin:09:sese}.

%
%
\begin{table*}[htb]
\caption{The two-loop self-energy
correction for the ground state of hydrogen-like ions, in units of $\Delta
E/[m\alpha^2(\Za)^4/\pi^2]$.
 \label{tab:sese}}
\begin{center}
\begin{tabular}{r......}
\hline\noalign{\smallskip}
$Z$  &  \multicolumn{1}{c}{LAL}
              &  \multicolumn{1}{c}{$F$ term}
                             &  \multicolumn{1}{c}{$P$ term}
                                      &  \multicolumn{1}{c}{$M$ term}
                                               &  \multicolumn{1}{c}{Total}
                                               &  \multicolumn{1}{c}{2005
                                                 results
                                                 \cite{yerokhin:05:sese}}\\
\noalign{\smallskip}\hline\noalign{\smallskip}
10 &  -0.3x58 &  822.x138\,(5) &  -721.x311\,(6) & -100.x297\,(35) &   0.17x2\,(36) & 0.25x\,(16) \\
12 &  -0.4x17 &  519.x603\,(2) &  -439.x065\,(6) &  -80.x117\,(38) &   0.00x4\,(38) & \\
15 &  -0.4x95 &  292.x901\,(2) &  -235.x211\,(4) &  -57.x406\,(11) &  -0.21x2\,(12) & -0.16x4\,(85) \\
17 &  -0.5x41 &  211.x052\,(1) &  -164.x280\,(3) &  -46.x567\,(9)  &  -0.33x6\,(10) & \\
20 &  -0.6x02 &  136.x909\,(1) &  -102.x029\,(2) &  -34.x780\,(4)  &  -0.50x1\,(5)  & -0.48x1\,(58)\\
25 &  -0.6x86 &   74.x501\,(1) &   -51.x982\,(2) &  -22.x560\,(6)  &  -0.72x8\,(6)  & \\
30 &  -0.7x56 &   44.x728\,(1) &   -29.x414\,(3) &  -15.x468\,(3)  &  -0.91x0\,(5)  & -0.90x3\,(26) \\
\noalign{\smallskip}\hline
\end{tabular}
\end{center}
\end{table*}

Table~\ref{tab:sese} summarizes our results for the total two-loop self-energy
correction for the ground state of hydrogen-like ions with the nuclear charge
$Z = 10-30$. We observe that
the calculational errors of the $P$ term do
not influence significantly the errors of the total results, as
the main uncertainty is now
delivered by the $M$ term. This uncertainty originates both from the
dependence of the results on the number of integration points and
from the termination of the partial-wave expansions. Since there are two
independent partial-wave expansion parameters in the $M$ term (see
Ref.~\cite{yerokhin:03:epjd} for details), the number of expansion terms
grows drastically as the cutoff parameter is increased. Because of this,
significant extension of the partial-wave summations looks prohibitively
expensive at present.

The two-loop self-energy correction for the ground
state of hydrogen-like atoms can be conveniently represented
in the following form, separating out the known terms of
the $\Za$ expansion,
\begin{align}
\label{G50} \Delta E&\ = m \left(\frac{\alpha}{\pi}\right)^2
(Z\alpha)^4\,
 \Bigl[
B_{40}+
(Z\alpha)\, G_{50}(Z)\Bigr] \,,
\end{align}
and
\begin{align} \label{G60}
G_{50}(Z) &\,=
B_{50}+(Z\alpha) \Bigl\{
  \ln^3[(Z\alpha)^{-2}]\, B_{63}
\nonumber \\ &
 +\ln^2[(Z\alpha)^{-2}]\,B_{62}
+  \ln[(Z\alpha)^{-2}]\,B_{61} + G_{60}(Z) \Bigr\}
\,,
\end{align}
where $B_{ij}$ are the expansion coefficients
with the first index corresponding to the power of $\Za$ and
the second index, to the power of logarithm, and $G_{ij}(Z)$ are the functions
incorporating the corresponding $B_{ij}$ and all higher orders in $\Za$,
$G_{ij}(Z) = B_{ij}+ \Za \,(\ldots)\,$.
The results for the expansion coefficients (see
Refs.~\cite{mohr:08:rmp,pachucki:01:pra,pachucki:03:prl,czarnecki:05:prl,%
jentschura:05:sese} and references
therein) are: $B_{40} = 1.409244$, $B_{50}= -24.2668(31)$, $B_{63}= -8/27$,
$B_{62}= 16/27-(16/9) \ln 2$, $B_{61} = 48.388913$, and $B_{60} = -61.6(9.2)$.

\begin{figure*}
\begin{center}\includegraphics[width=\textwidth]{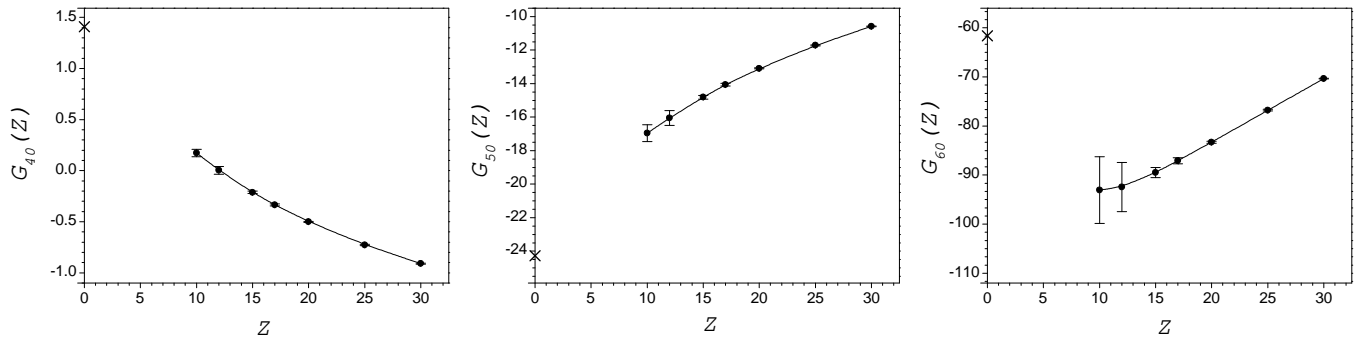}\end{center}%
\caption{\label{fig:ho}
The two-loop self-energy correction.
$G_{40}(Z) = \Delta E/[m\alpha^2(\Za)^4/\pi^2]$, $G_{50}(Z)$ and
$G_{60}(Z)$ are defined by Eqs.~(\ref{G50}) and (\ref{G60}).
The cross on the $y$-axis indicates the analytical results, $G_{40}(0) =
B_{40}$, $G_{50}(0) = B_{50}$, and $G_{60}(0) = B_{60}$.
}%
\end{figure*}

The functions $G_{ij}(Z)$ inferred from our numerical data are plotted in
Fig.~\ref{fig:ho}. The visual agreement of our 
results with the analytical values of the expansion coefficients is very good
for $B_{40}$ and $B_{50}$, but not exactly satisfactory for $B_{60}$.
In order to produce a clearer statement, we need to extrapolate our data
towards low values of $Z$. For this we use a variant of the procedure first
employed in Ref.~\cite{mohr:75:prl}.  The extrapolation
towards the required value of $Z=Z_0$ ($=0$ and $1$ in our case)
is performed in two steps. First, we apply an
(exact) linear fit to each pair of two consecutive points from our data set
and store the resulting values at $Z=Z_0$. Second, we perform a global parabolic
least-squares fit to the set of data obtained on the first step
and take the fitted value at
$Z=Z_0$ as a final result. Similar procedure applied to the function $G_{50}$
reproduces the analytical result for the coefficient $B_{50}$ with the accuracy
of about 1\%. For comparison, a global polynomial fit yields a result
accurate within 5\% only.

When applied to the remainder function
$G_{60}(Z)$, the extrapolation procedure described above gives
\begin{align} \label{ho}
G_{60}(Z=0) &\ = -84\,(15)\,, \\
G_{60}(Z=1) &\ = -86\,(15)\,.
\label{hop}
\end{align}
The extrapolated value for $Z=1$ is higher than but marginally
consistent with the 2005 result of $-127\,(42)$
\cite{yerokhin:05:sese}. The shift of the central value is due to two
reasons. First, the analytical result for the $B_{61}$ coefficient was recently changed
by $\delta B_{61}=-1.4494\ldots$ \cite{jentschura:05:sese}, thus
pushing the remainder function higher up. Second, the improved numerical
accuracy of the present calculation and the increased number of values of $Z$
studied allowed us to identify the upward trend in the numerical data.

The shift of the extrapolated values for $G_{60}$ 
significantly reduced the disagreement 
with the analytical calculation \cite{pachucki:03:prl} reported in Ref.~\cite{yerokhin:05:sese}.
The present value of $G_{60}(0) = -84(15)$ is consistent (but still not in perfect
agreement) with the analytical result of $B_{60} = -62(9)$ .

To complete our analysis of the higher-order two-loop effects in
hydrogen, we combine the result for the two-loop self-energy
correction obtained in this work [Eq.~(\ref{hop})] with the corresponding
contribution induced by the two-loop diagrams with the closed fermion loops
reported in Ref.~\cite{yerokhin:08:twoloop}. So, our estimate of the {\em total}
two-loop (nonlogarithmic) contribution to order $m\alpha^2(\Za)^6$ for the
ground state of hydrogen is
\begin{equation}
G_{60}(Z=1, \mbox{\rm total}) = -86\,(15)-15\,(2) = -101\,(15)\,.
\end{equation}
The numerical contribution of this effect is $-10.2(1.5)$~kHz,
which is much larger than the error of the experimental
determination of the $1S-2S$ transition frequency in hydrogen
\cite{fischer:04} (34~Hz) and comparable with the experimental
errors for the $2S-12D$ transitions
\cite{schwob:99} (7~kHz).

To conclude, in the present investigation, we described in detail
the technique of evaluation of Feynman diagrams in the mixed
coordinate-momentum representation based on the analytical representation of the
bound electron propagators in terms of the Whittaker functions. This technique
allowed us to significantly improve the accuracy of the numerical evaluation
of the part of the two-loop self-energy correction conventionally termed as
the $P$ term. The all-order (in the parameter $\Za$) results are reported for
the two-loop self-energy correction for the ground state of hydrogen-like ions
with with the nuclear charge number $Z=10-30$.
The higher-order (in $\Za$) remainder function is inferred from the numerical
results and extrapolated towards $Z=0$ and $1$. The extrapolated value
is in marginal agreement with the
analytical result obtained within the perturbative approach.

The work presented in this paper was supported by RFBR (grant No.~10-02-00150-a).


\appendix
\begin{widetext}
\section{Dirac Coulomb Green function in the coordinate-momentum representation}
\label{app:DCGF}

The Dirac Coulomb Green function is commonly written in coordinate space
as an expansion in the relativistic angular momentum parameter $\kappa$ 
\cite{wichmann:56,mohr:74:a,mohr:98},
\begin{align}
 G(E,\bfx_1,\bfx_2) = \sum_{\kappa\mu} \left(
        \begin{array}{ll}
        G_{\kappa}^{11}(E,x_1,x_2)\,\chi_{\kappa\mu}(\hx_1)\,\chi_{\kappa\mu}^{\dag}(\hx_2)
      & G_{\kappa}^{12}(E,x_1,x_2)\,(-i)\,\chi_{\kappa\mu}(\hx_1)\,\chi_{-\kappa\mu}^{\dag}(\hx_2)    \\
        G_{\kappa}^{21}(E,x_1,x_2)\,i\,\chi_{-\kappa\mu}(\hx_1)\,\chi_{\kappa\mu}^{\dag}(\hx_2)
      & G_{\kappa}^{22}(E,x_1,x_2)\,\chi_{-\kappa\mu}(\hx_1)\,\chi_{-\kappa\mu}^{\dag}(\hx_2)    \\
        \end{array}
\right)\,,
\end{align}
\end{widetext}
where $\chi_{\kappa \mu}(\hx)$ are the Dirac spin-angular spinors
\cite{rose:61}
and $\hx = \bfx/|\bfx|$.
The $2\times2$ matrix of the radial components $G_{\kappa}^{ij}$ is
referred to as the radial Green function and denoted as $G_{\kappa}$.
The radial Green function can be expressed in terms
of the two-component solutions of the radial Dirac equation
regular at the origin $\left(\phi_{\kappa}^{0}\right)$
and the infinity $\left(\phi_{\kappa}^{\infty}\right)$
as follows
\begin{align}\label{gr01}
 G_{\kappa}(E,x_1,x_2) = &\,
 -\phi_{\kappa}^{\infty}(E,x_1)\,\phi_{\kappa}^{0^T}(E,x_2)\,\theta(x_1-x_2)
\nonumber \\ &
 -\phi_{\kappa}^{0}(E,x_1)\,\phi_{\kappa}^{{\infty}^T}(E,x_2)\,\theta(x_2-x_1)\,.
\end{align}
The upper and the lower components of the functions $\phi_{\kappa}^{0}$ and
$\phi_{\kappa}^{\infty}$ will be denoted by subscripts "+" and "-", respectively.
They are given by
\begin{align} \label{Green2}
\phi_{\kappa,\pm}^{0}(E,x) =&\, \Delta_{\kappa}^{-1/2}\,
        \frac{\sqrt{1\pm E}}{x^{3/2}}
        \Bigl[ (\lambda-\nu)\,M_{\nu-\frac12,\lambda}(2cx)
\nonumber \\ &
        \mp \left(\kappa-\frac{\alpha Z}{c}\right)\,M_{\nu+\frac12,\lambda}(2cx) \Bigr]
    \ ,
\end{align}
\begin{align}
\phi_{\kappa,\pm}^{\infty}(E,x) =&\, \Delta_{\kappa}^{-1/2}\,
        \frac{\sqrt{1\pm E}}{x^{3/2}}
        \Bigl[ \left(\kappa+\frac{\alpha Z}{c}\right)\,W_{\nu-\frac12,\lambda}(2cx)
\nonumber \\ &
        \pm W_{\nu+\frac12,\lambda}(2cx) \Bigr] \ ,
\label{Green2a}
\end{align}
where $\Delta_{\kappa} = 4c^2\, \Gamma(1+2\lambda)/\Gamma(\lambda-\nu)$,
$c=\sqrt{1-E^2}$ defined so that $\Re(c)>0$, $\lambda=\sqrt{\kappa^2-(Z\alpha)^2}$,
$\nu=Z\alpha E/c$, and $M_{\alpha,\beta}$ and
$W_{\alpha,\beta}$ are the Whittaker functions of the first and the second
kind, respectively.

The Dirac Coulomb Green function in the coordinate-momentum representation is
obtained from the above formulas by the Fourier transform over one of the radial
arguments. Let us consider the transform over, e.g., the second radial
argument,
\begin{align}
 G(E,\bfx_1,\bfp_2) = \int d\bfx_2\, e^{i\bfp_2\cdot\bfx_2}\,G(E,\bfx_1,\bfx_2)\,.
\end{align}
Its partial-wave expansion takes the form
\begin{widetext}
\begin{align}\label{gr0}
 G(E,\bfx_1,\bfp_2) = \sum_{\kappa\mu} i^l\left(
        \begin{array}{ll}
        G_{\kappa}^{11}(E,x_1,p_2)\,\chi_{\kappa\mu}(\hx_1)\,\chi_{\kappa\mu}^{\dag}(\hp_2)
      & G_{\kappa}^{12}(E,x_1,p_2)\,\chi_{\kappa\mu}(\hx_1)\,\chi_{-\kappa\mu}^{\dag}(\hp_2)    \\
        G_{\kappa}^{21}(E,x_1,p_2)\,i\,\chi_{-\kappa\mu}(\hx_1)\,\chi_{\kappa\mu}^{\dag}(\hp_2)
      & G_{\kappa}^{22}(E,x_1,p_2)\,i\,\chi_{-\kappa\mu}(\hx_1)\,\chi_{-\kappa\mu}^{\dag}(\hp_2)    \\
        \end{array}
\right)\,,
\end{align}
with the radial part given by the
following matrix
\begin{align}\label{gr1}
 G_{\kappa}(E,x_1,p_2) = 4\pi\, \int_0^{\infty}dx_2\,x_2^2\,
 \left(
        \begin{array}{ll}
          j_l(p_2x_2)\,G^{11}_{\kappa}(E,x_1,x_2) &
           -\frac{\kappa}{|\kappa|}\,j_{\overline{l}}(p_2x_2)\,G^{12}_{\kappa}(E,x_1,x_2) \\
          j_l(p_2x_2)\,G^{21}_{\kappa}(E,x_1,x_2) &
           -\frac{\kappa}{|\kappa|}\,j_{\overline{l}}(p_2x_2)\,G^{22}_{\kappa}(E,x_1,x_2) \\
        \end{array}
 \right)\,,
\end{align}
where 
$l = |\kappa+1/2|-1/2$ and $\overline{l} = |\kappa-1/2|-1/2$. Using Eq.~(\ref{gr01}), we obtain the
following representation for the radial Green function in the mixed
coordinate-momentum representation,
\end{widetext}

\begin{align}\label{gr2}
 G_{\kappa}(E,x_1,p_2)  = &\,
 -\phi_{\kappa}^{\infty}(E,x_1)\,\psi_{\kappa}^{0^T}(E,p_2;x_1)
\nonumber \\ &
 -\phi_{\kappa}^{0}(E,x_1)\,\psi_{\kappa}^{{\infty}^T}(E,p_2;x_1)\,,
\end{align}
where
\begin{align}\label{gr3}
\psi_{\kappa}^{0}(E,p;x_1) = &\, 4\pi\,
\int_0^{x_1}dx_2\,
  x_2^2\,
\nonumber \\ & \times
\left(
        \begin{array}{r}
   j_l(px_2)\,\phi^0_{\kappa,+}(E,x_2) \\[0.5em]
   -\frac{\kappa}{|\kappa|\,}j_{\overline{l}}(px_2)\,\phi^0_{\kappa,-}(E,x_2) \\
        \end{array}
\right)\,,
\end{align}
and
\begin{align}\label{gr4}
\psi_{\kappa}^{\infty}(E,p;x_1) = &\, 4\pi\,\int_{x_1}^{\infty} dx_2\,
  x_2^2\,
\nonumber \\ & \times
\left(
        \begin{array}{r}
   j_l(px_2)\,\phi^{\infty}_{\kappa,+}(E,x_2) \\[0.5em]
   -\frac{\kappa}{|\kappa|\,}j_{\overline{l}}(px_2)\,\phi^{\infty}_{\kappa,-}(E,x_2) \\
        \end{array}
\right)\,.
\end{align}

The integration over $x_2$ in the functions $\psi_{\kappa}^0$ and
$\psi_{\kappa}^{\infty}$ has to be
performed numerically. The problems here are that (i) the integration interval depends on
$x_1$ and (ii) the integrand contains the spherical Bessel function which
oscillates rapidly in the high-momenta region.
Clearly, a straightforward use of Eqs.~(\ref{gr3}) and (\ref{gr4}) in our calculations
would lead to a re-evaluation of the integral for each new value of $x_1$,
making the computation prohibitively expensive.
One can observe, however, that if the function $\psi_{\kappa}(E,p;x)$ is known
for a particular set of $E$, $p$, and $x$, then the evaluation of
$\psi_{\kappa}(E,p;x\pr)$ can be done by computing the Bessel
transform integral over the interval $(x,x\pr)$ only. So, introducing an ordered
radial grid $\left\{x_i\right\}$, one can store
the whole set of values $\left\{\psi_{\kappa}(E,p;x_i)\right\}$ by performing just one Bessel
transform over the interval $(0,\infty)$. This shows that for a fixed values of $E$ and $p$,
the integrations of the type $\int_0^{\infty} dx\, f(x)\, \psi_{\kappa}(E,p;x)$
can be performed without a recalculation of the
Bessel transform integral.

\section{Free Dirac Green function in the coordinate-momentum representation}
\label{app:freeDGF}

The free Dirac Green function $G^{(0)}$ is a much simpler object
than the Dirac Coulomb Green
function $G$ and is known in the closed analytical form as well as in the
partial-wave expansion form (see, e.g., Ref.~\cite{mohr:74:a}).
For the purposes of the present investigation, we employ the
coordinate-momentum representation and put $G^{(0)}$ into the form analogous to
Eqs.~(\ref{gr0}) and (\ref{gr1}). The simplest way to achieve this is to start with the
momentum representation of $G^{(0)}$, which has a particularly simple form,
\begin{align} \label{fgr1}
G^{(0)}(E,\bfp_1,\bfp_2) &\,= (2\pi)^3\, \frac{\delta^3(\bfp_1-
    \bfp_2)}{\gamma^0E - \bgamma\cdot\bfp_2-m}\gamma^0
\nonumber \\ &
 = (2\pi)^3\,
\frac{E+ \balpha\cdot\bfp_2+m\gamma^0}{E^2-\bfp_2^2-m^2}\,\delta^3(\bfp_1-\bfp_2)\,,
\end{align}
where $\balpha = \gamma^0\bgamma$.
Using the completeness of the angular-momentum spinors $\chi_{\kappa \mu}$,
\begin{equation} 
  \sum_{\kappa \mu} \chi_{\kappa \mu}(\hp_1)\, \chi^{\dag}_{\kappa
    \mu}(\hp_2) =
   I\, \delta(\hp_1-\hp_2)\,,
\end{equation}
and the identity
$  (\bsigma\cdot\hp)\, \chi_{\kappa \mu}(\hp) = -\chi_{-\kappa \mu}(\hp)\,,$
we cast Eq.~(\ref{fgr1}) into the partial-wave expansion form similar to that
for the Dirac Coulomb Green function,

\begin{widetext}
\begin{align} \label{fgr2}
G^{(0)}(E,\bfp_1,\bfp_2) = (2\pi)^3\,
\frac{\frac1{p_2^2}\,\delta(p_1-p_2)}{E^2-p_2^2-m^2}\,
 \sum_{\kappa \mu}
\left(
        \begin{array}{rr}
          (E+m)\,\chi_{\kappa\mu}(\hp_1)\,\chi_{\kappa\mu}^{\dag}(\hp_2)
      & -p_2\,\chi_{\kappa\mu}(\hp_1)\,\chi_{-\kappa\mu}^{\dag}(\hp_2)    \\
        -p_2\,\chi_{-\kappa\mu}(\hp_1)\,\chi_{\kappa\mu}^{\dag}(\hp_2)
      & (E-m)\,\chi_{-\kappa\mu}(\hp_1)\,\chi_{-\kappa\mu}^{\dag}(\hp_2)    \\
        \end{array}
\right)\,,
\end{align}
where $p_i = |\bfp_i|$.
The coordinate-momentum representation of $G^{(0)}$ is obtained 
by the Fourier transform of the above expression 
over the first radial argument,
\begin{align}
 G^{(0)}(E,\bfx_1,\bfp_2) = \int \frac{d\bfp_1}{(2\pi)^3}\,
 e^{i\bfp_1\cdot\bfx_1}
      \,G^{(0)}(E,\bfp_1,\bfp_2)\,.
\end{align}
After performing the integration over $\bfp_1$,
the free Dirac Green function is written in the form of
Eq.~(\ref{gr0}), with the radial part given by
\begin{align}\label{fgr3}
 G^{(0)}_{\kappa}(E,x_1,p_2) = \frac{4\pi}{E^2-p_2^2-m^2}\,
 \left(
        \begin{array}{rr}
          (E+m)\,j_l(p_2x_1) &
           -p_2\,j_{l}(p_2x_1) \\
          \frac{\kappa}{|\kappa|}\,p_2\,j_{\overline{l}}(p_2x_1) &
          - \frac{\kappa}{|\kappa|}\,(E-m)\,j_{\overline{l}}(p_2x_1) \\
        \end{array}
 \right)\,.
\end{align}
\end{widetext}

\section{Angular factors}
\label{app:angular}

In this section we address the factors
$t_{\kappa_n,\kappa_a}$ and $s^k_{\kappa_n,\kappa_a}$,
which are defined by Eqs.~(\ref{pef6})-(\ref{pef6c}).
Inserting the explicit definitions of the angular-momentum spinors
in these formulas,
averaging over the momentum projections of the reference state,   and
calculating the sums of the Clebsch-Gordan coefficients, we arrive at the
following results,
\begin{align}    \label{C1}
t_{\kappa_n,\kappa_a}(J) = &\
\frac{(-1)^{j_a+1/2}}{\sqrt{4\pi}} \frac{\Pi_{j_n}}{\Pi_{j_a}}\,
  \SixJ{j_a}{j_n}{J}{l_n}{l_a}{1/2}\,
\nonumber \\ & \times
 \sum_M (-1)^M Y_{JM}(\hq)\, Y^{J\,-M}_{l_n l_a}(\hp_1,\hp_2)\,,
\end{align}
\begin{align}
s_{\kappa_n,\kappa_a}^{\sigma}(JL)  = &\
 (-1)^{l_n+L}\,\sqrt{\frac{6}{4\pi}}\, \frac{\Pi_{j_nJJ}}{\Pi_{j_aL}}\,
\NineJ{j_a}{j_n}{J}{1/2}{1/2}{1}{l_a}{l_n}{L}\,
\nonumber \\ & \times
 \sum_M (-1)^M Y_{LM}(\hq)\, Y^{L\,-M}_{l_n l_a}(\hp_1,\hp_2)\,,
\end{align}
\begin{align}    \label{C3}
s_{\kappa_n,\kappa_a}^{p_1}(JL)  = &\
\frac{(-1)^{j_a+1/2}}{\sqrt{3}}\, \frac{\Pi_{j_n}}{\Pi_{j_a}}\,
  \SixJ{j_a}{j_n}{J}{l_n}{l_a}{1/2}\,
\nonumber \\ & \times
 \sum_M (-1)^M Y_{L1}^{JM}(\hq,\hp_1)\, Y^{J\,-M}_{l_n l_a}(\hp_1,\hp_2)\,,
\end{align}
\begin{align}    \label{C4}
s_{\kappa_n,\kappa_a}^{p_2}(JL)  = &\
\frac{(-1)^{j_a+1/2}}{\sqrt{3}}\, \frac{\Pi_{j_n}}{\Pi_{j_a}}\,
  \SixJ{j_a}{j_n}{J}{l_n}{l_a}{1/2}\,
\nonumber \\ & \times
 \sum_M (-1)^M Y_{L1}^{JM}(\hq,\hp_2)\, Y^{J\,-M}_{l_n l_a}(\hp_1,\hp_2)\,,
\end{align}
where $\Pi_{j_1j_2\ldots} = \sqrt{(2j_1+1)(2j_2+1)\ldots}$ and $Y^{JM}_{l_1
  l_2}(\hp_1,\hp_2)$ are the bipolar spherical harmonics \cite{varshalovich}.

With help of formulas from the book \cite{varshalovich}, it is possible to
obtain explicit results for the angular factors
$t_{\kappa_n,\kappa_a}$ and $s^k_{\kappa_n,\kappa_a}$, which
are functions of $p_1 = |\bfp_1|$, $p_2 = |\bfp_2|$, and $q = |\bfp_1-\bfp_2|$
only. However, the resulting formulas turn out to be
rather lengthy and not very convenient for numerical evaluation
as they become numerically unstable for $q\to 0$. Because of
this, we prefer to evaluate Eqs.~(\ref{C1})-(\ref{C4})
numerically, after some simplifications that exploit the fact that the result does not
depend on any angles except for $\hp_1\cdot\hp_2$. Namely, we set the
azimuthal spherical coordinate of $\hp_1$ and $\hp_2$ to zero ($\phi_1 = \phi_2
= 0$) and direct $\hp_1$ along the $z$ axis ($\theta_1 = 0$).

\end{document}